\documentclass[twocolumn,twoside,slac_two]{revtex4}
\usepackage{graphicx}
\usepackage{fancyhdr}
\usepackage{graphicx,axodraw,subfigure}

\def\nn{\nonumber}

\def\beq{\begin{equation}}
\def\eeq{\end{equation}}
\def\bea{\begin{eqnarray}}
\def\eea{\end{eqnarray}}

\begin{document}
\title{CP violation in $\tau$ decays}
%
\author{Daiji Kimura$^1$, Kang Young Lee$^2$,
 Takuya Morozumi$^1$,
and Keita Nakagawa$^1$}
\affiliation{$^1$ Graduate School of Science, Hiroshima University, 
Higashi-Hiroshima, 739-8526, Japan \\
$^2$ Division of Quantum Phases \& Devices, School of Physics, Konkuk University, Seoul 143-701, Korea}
\begin{abstract}
CP violation of the charged lepton sector has not been 
found yet. $\tau$ is a unique charged lepton which can decay
into hadrons. In this talk, we argue what kind of new physics
can be studied with CP violation of $\tau$ lepton decays.
We also disucuss how to handle hadronic final state
interactions for a prediction of the direct CP violation. 
\end{abstract}
\maketitle
\thispagestyle{fancy}
\section{Introduction}
CP violation of the lepton sector has not been 
found yet.  CP violation of the neutrino sector
will be explored by measuring the asymmetry
of the oscillation probabilities:
$\nu_{\mu} \to \nu_{e}$ and $\bar{\nu}_{\mu} \to \bar{\nu}_e$.
The CP violation requires the flavor violation and non-degeneracy
of the masses for the charged leptons and neutrinos.
Yet another type of CP violation of the lepton sector 
can be explored using the $\tau$ lepton which can be produced
abundantly in B factories and Super flavor factories.

First let us summarize the present status of CP and T 
violation of $\tau$ lepton.
The experimental bound on CP(T) violation related to $\tau$
lepton are derived from measurements
of the electric dipole moment (EDM) and CP violation
of hadronic 
$\tau$ decays. 
The EDM of $\tau$ lepton is given by the operator 
\bea
i e d_{\rm \tau} \overline{\tau} \sigma_{\mu \nu} \gamma_5 
 \tau F^{\mu \nu}.
\label{edm}
\eea
In the non-relativistic limit, using the two component spinor
$\phi$, the operator in Eq.~(\ref{edm}) is reduced to 
\bea
2 e d_{\rm \tau} \phi^{\dagger} {\bf S} \phi \cdot {\bf E},
\eea
with ${\bf S}=\frac{\bf{\sigma}}{2}$.
The non-zero coefficient $d_{\rm \tau}$ implies T violation
because the operator is a T odd operator,
\bea
{\bf E} \rightarrow -{\bf E}, \quad
\phi^{\dagger} {\bf S} \phi \to \phi^{\dagger} {\bf S} \phi.
\eea
The recent experimental bounds on electric dipole moment
(edm) of $\tau$ leptons are
summarized as,
\bea
&& -0.22 \times 10^{-16} \le {\rm Re} (e \ d_{\tau}) \le 0.45
 \times 10^{-16}({\rm e} \ {\rm cm}), \nn \\
&& -0.25 \times 10^{-16}
\le {\rm Im}(e \ d_{\tau}) \le 0.008 \times 10^{-16}.
\label{bound}
\eea
The second example of CP violation is in
$\tau$ hadronic decays.
There is "known" CP violation of 
the semileptonic $\tau$ decays even within the
standard model ~\cite{Bigi:2005ts, Calderon},
\bea
\frac{\Gamma[\tau^- \to K_s \pi^- \nu]}
{\Gamma[\tau^+ \to K_s \pi^+ \overline{\nu}]} &=&\frac{
|q|^2}{|p|^2} 
\frac{\Gamma[\tau^- \to \overline{K^0} \pi^- \nu]}{\Gamma[\tau^+
\to K^0 \pi^+ \overline{\nu}]},\nn \\
&=& \frac{1-A_L}{1+A_L},
\eea
where in the second line, we assume
$\frac{\Gamma[\tau^- \to \overline{K^0} \pi^- \nu]}{\Gamma[\tau^+
\to K^0 \pi^+ \overline{\nu}]}=1$ which is valid within the
tree level approximation of the standard model.
$A_L $ is CP violation of the charge asymmetry of 
$K_L \to \pi^- l^+
\nu_l$ and $K_L \to \pi^+ l^- \bar{\nu}_l$ with $l=e, \mu$
\bea
A_L&=&\frac{\Gamma[K_L \to  \pi^- l^+ \nu_l]-
\Gamma[K_L \to  \pi^+ l^- \bar{\nu}_l]}
{\Gamma[K_L \to  \pi^- l^+ \nu_l]+
\Gamma[K_L \to  \pi^+ l^- \bar{\nu}_l]},\nn \\
&=& \frac{|p|^2-|q|^2}{|p|^2+|q|^2}=(3.32 \pm 0.06) \times 10^{-3},
\label{AL}
\eea
where $p$ and $q$ are the mixing amplitudes of  
$K^0$ and $\bar{K^0}$ in the mass eigenstates $K_{S,L}$
\bea
|K_S \rangle&=&p|K^0(\overline{s} d)\rangle + q |\overline{K^0}
(s \overline{d})\rangle, \nn \\
|K_L \rangle&=& p|K^0(\overline{s} d)\rangle -q |\overline{K^0}
(s \overline{d})\rangle. 
\eea
The first limit of CP violation of $\tau$ decay is set by CLEO
~\cite{Bonvicini:2001xz}.
Assuming the parametrization for the decay $\tau^{-} \to K^{-}
\pi^0 \nu$,  
\bea
A(\tau^- \to K^- \pi^0 \nu_{\tau})&=&\bar{u} \gamma_{\mu}(1-\gamma_5)
u_{\tau} \times \nn \\
&& f_V (-q^{\mu}+ \frac{\Delta_{K \pi}}{Q^2} Q^{\mu}) \nn \\
&+& \Lambda \bar{u} (1+\gamma_5)u_{\tau} 
f_s M.
\label{CLEO}
\eea
where $q=p_K-p_\pi, Q=p_K+p_{\pi}$ and $M=1$(GeV).
The bound on the parameter of CP violation is obtained as
~\cite{Bonvicini:2001xz},
\bea
-0.172 <{\rm Im} \Lambda< 0.067.
\eea
In the present talk, 
we give our predictions for the CP violation of
$\tau \to K \pi \nu $ decays in a two Higgs
doublet model.
The results including the $K \eta$ and $K \eta'$
and the improved form factors are given in 
~\cite{Kimura:2008gh}. 
\section{What kind of CP violation might manifest itself
in $\tau$ decay ? }
We first give an example of the beyond the standard model
which gives rise to the CP violation
in $\tau$ hadronic decays ~\cite{Kuhn:1996dv, Tsai:1996ps,
Choi:1998yx}.
The direct CP violation of $\tau$ decay may appear when 
there are two or more interfering amplitudes with
different weak phases and the strong phases.
In $\tau^- \to K^- \pi \nu$ and its CP conjugate process
$\tau^+ \to K^+ \pi \bar{\nu}$, within the standard model,
the charged current interaction leads to the decays as,
\bea
&& \tau^- \to \nu_{\tau} W^{- \ast} \to \nu_{\tau} \bar{u} s \to 
\nu_{\tau} K^- \pi^0, \nn \\
&& \tau^+ \to \bar{\nu}_{\tau} W^{+ \ast} 
\to \bar{\nu}_{\tau}  \bar{s} u \to \bar{\nu}_{\tau} K^+ \pi^0. 
\eea  
In the two Higgs 
doublet model, the charged Higgs interaction
may also generate the amplitude,
\bea
\tau_R^- \to \nu_{i} H^{- \ast} \to \nu_{i}  s_R \overline{u_L} 
\to \nu_i  K^- \pi^0,
\label{chargedhiggs}
\eea
where $i= e, \mu$ and $\tau$. When the flavor of the
neutrino is $\tau-$flavored, the charged Higgs contribution
may interfere with the contribution of the standard model.
The charged Higgs contributes to the angular momentum
$L=0$ state (s wave) of the hadronic (K $\pi$) system
in the hadronic rest frame and 
the charged current due to $W$ boson
interaction
contributes to $L=1,0$ state. The strong phase
due to the final state interaction for the s wave
of $K$ and $\pi$ is 
different from the phase shift of the p wave. Moreover,
the CP phase of the charged Higgs
contribution may be also differerent from the weak phase 
$V_{us}$ of the standard model contribution. Therefore 
one may expect
the direct CP violation in the interference of the amplitudes
of the p wave and the s wave of $K$ and $\pi$ system.
On the flavor of neutrino in Eq.(\ref{chargedhiggs}),
by ignoring the mass term of the neutrino compared with
the $\tau$ lepton mass, 
only in the case that the flavor of the 
neutrino is $\tau$ flavored, the interference occurs.
For the other flavor case which corresponds to $\nu_e$ and $\nu_\mu$, 
the interference term in the charged current contribution
is absent. Therefore, the direct CP violation
may not be expected from the flavor changing case in 
Eq.(\ref{chargedhiggs}).
\section{CP violation of $\tau$ decay, edm and 
flavor changing neutral current (FCNC).}
\begin{figure}[htbp]
\begin{center}
\includegraphics[width=8cm]{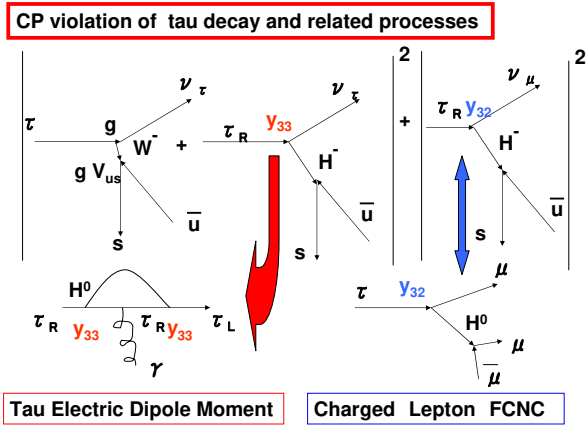} 
\end{center}
\caption{CP violation and FCNC processes related to the charged
Higgs and neutral Higgs boson exchanged processes.}
\label{fig1}
\end{figure}
In Fig.(\ref{fig1}), it is shown how the
direct CP violation of $\tau$ hadronic decay
and edm of the $\tau$ lepton is related to each other
in the two Higgs doublet model. 
We also show, in the same model, how
the flavor changing interaction in 
charged current mediated by the charged Higgs boson 
exchange
is related to the flavor changing neutral current (FCNC)
mediated by the neutral Higgs boson.
Since the charged Higgs boson forms the SU(2) doublet with the
neutral Higgs boson, we can expect the CP violation and
the flavor changing 
in Eq.(\ref{chargedhiggs}) also
contributes to the CP violation and flavor changing
in the neutral Higgs sector, namely,
\bea
(\tau_R \to H^- \nu_{iL}) \leftrightarrow 
(\tau_R \to H^0 l_{iL}).     
\eea
where $H^0$ is a linear combination of the three
mass eigenstates of the neutral Higgs bosons. 
For $i=\tau$, the flavor diagonal coupling 
of $\tau$ lepton and neutral/charged Higgs boson denoted by $y_{33}$
is CP violating
and it contributes to the $\tau$ electric dipole moment
as shown in Fig.~(\ref{fig1}). For the flavor changing 
coupling $i=\mu$ denoted by $y_{32}$ in Fig.~(\ref{fig1}),
it also leads to the flavor changing 
neutral current process in the charged lepton such as $\tau \to 
\mu \bar{\mu} \mu $ mediated by the neutral  
Higgs boson exchange. Therefore, one may constrain the
flavor changing coupling of Eq.~(\ref{chargedhiggs}) from
the charged lepton FCNC process such as $\tau \to l_i l_j^+ l_j^-$
($l_i, l_j =\mu, e$) while CP violation of the flavor diagonal
coupling can be constrained from edm of $\tau$ lepton.
\section{CP violation measurement of the unpolarized $\tau$ decay}
Now we turn to the measurement of the CP violation.
The issue has been disucussed in \cite{Kuhn:1996dv}.
In order to extract the interference term of $L=0$ and $L=1$
states of kaon and pion, we must measure the angular 
distribution. 
\begin{figure}[thbp]
\begin{center}
\includegraphics[width=7cm]{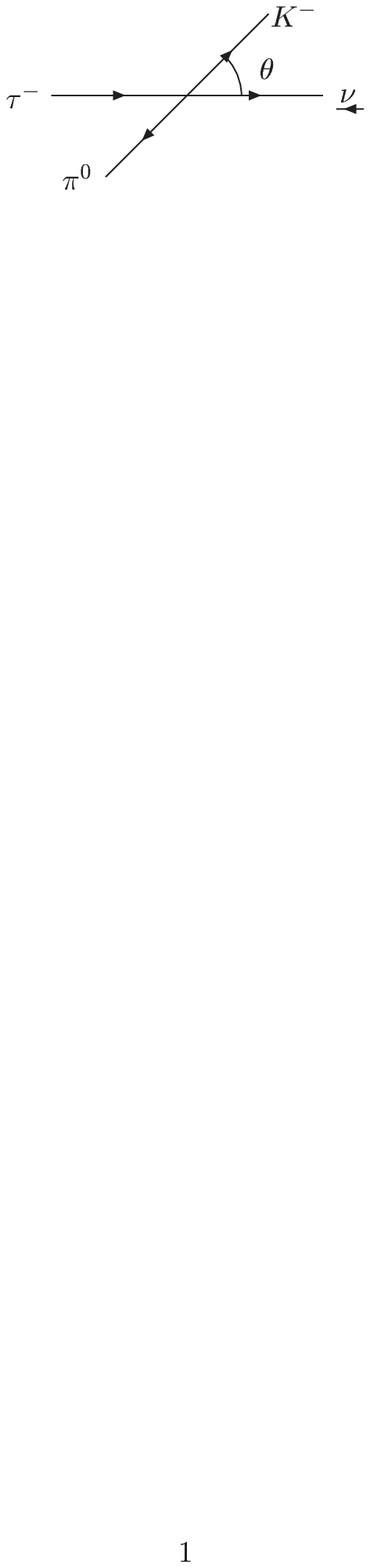}
\end{center}
\caption{The angle $\theta$ defined at CM frame
of K and $\pi$.}
\label{theta}
\end{figure}
The angular distribution within the
standard model is given below, which can be also read
off from Eq.(10) of \cite{Kuhn:1996dv} by replacing 
$\beta$ in Eq.(29) with $\theta$ where $\cos \theta=
{\bf n_{\tau}} \cdot {\bf n_k}$. Here, ${\bf n_k}$ and
${\bf n_\tau}$
are the
directions of the kaon and $\tau$ lepton in the hadronic rest frame
respectively as shown in Fig.~(\ref{theta}),
\bea
&& \frac{d^2\Gamma}{d\sqrt{s} d\cos \theta}=\frac{G_F^2 |V_{us}|^2}{2^5 \pi^3}
\frac{(m_{\tau}^2-s)^2}{m_{\tau}^3} l(s) \nn \\
&& \left( (\frac{m_\tau^2}{s} \cos^2 \theta + \sin^2 \theta) l(s)^2 
|F(s)|^2 +\frac{m_{\tau}^2}{4} |F_s(s)|^2 \right. \nn \\
&& \left. -\frac{m_\tau^2}{\sqrt{s}} l(s) \cos \theta
{\rm Re}(F F_s^{\ast})  \right),  
\eea
where $l(s)$ is the three momentum of kaon in the hadronic
rest frame. $\sqrt{s}$ is the invariant mass
for hadrons. 
$F$ and $F_S$ are the vector and the scalar form factors 
defined below.
\bea
&& \langle K^+(p_K) P(p_P)|\bar{u} \gamma_{\mu} s|0 \rangle
=F(Q^2)q^{\mu} \nn \\
&+& \left(F_S(Q^2)- 
\frac{\Delta_{K P}}{Q^2} F(Q^2)\right) Q^{\mu},  
\label{formfactors}
\eea
with $Q^{\mu}=(p_K+p_\pi)^{\mu}$ and $\Delta_{KP}=m_K^2-m_\pi^2$.
From the angular distribution,
one can define the forward and the backward
asymmetry \cite{Beldjoudi:1994hi},
\bea
&& A_{\rm FB}(s)=
\frac{{\frac{d \Gamma}{d\sqrt{s}}}\bigg|_{\cos \theta >0}  -
{\frac{d \Gamma}{d \sqrt{s}}}\bigg|_{\cos \theta <0}}
{\frac{d {\rm \Gamma}}{d \sqrt{s}}}, \nn \\
&&{\frac{d \Gamma}{d\sqrt{s}}}\bigg|_{\cos \theta >0}  -
{\frac{d \Gamma}{d \sqrt{s}}}\bigg|_{\cos \theta <0} \nn \\
&=& -\frac{G_F^2 |V_{us}|^2} {2^5 \pi^3}
\frac{(m_{\tau}^2-s)^2}{m_{\tau}^3} 
\frac{m_{\tau}^2}{\sqrt{s}}
l(s)^2 {\rm Re.}{(F F_S^{\ast})},
\label{FB}
\eea
As we can see from Eq.~(\ref{FB}), the
forward and backward asymmetry defined at the hadronic rest frame
is 
the difference of the numbers of the events for the
kaon scattered 
into the forward direction and the backward direction
with respect to the incoming
$\tau$. Note that the asymmetry
is proportional to the interference term of the vector
and the scalar form factors.
\section{The scalar and vector form factors from the
chiral Lagrangian}
\begin{figure}[htbp]
\begin{center}
\begin{tabular}{cc}
 $|F|$ & ${\rm Re} F$ \\
\resizebox{35mm}{!}{\includegraphics{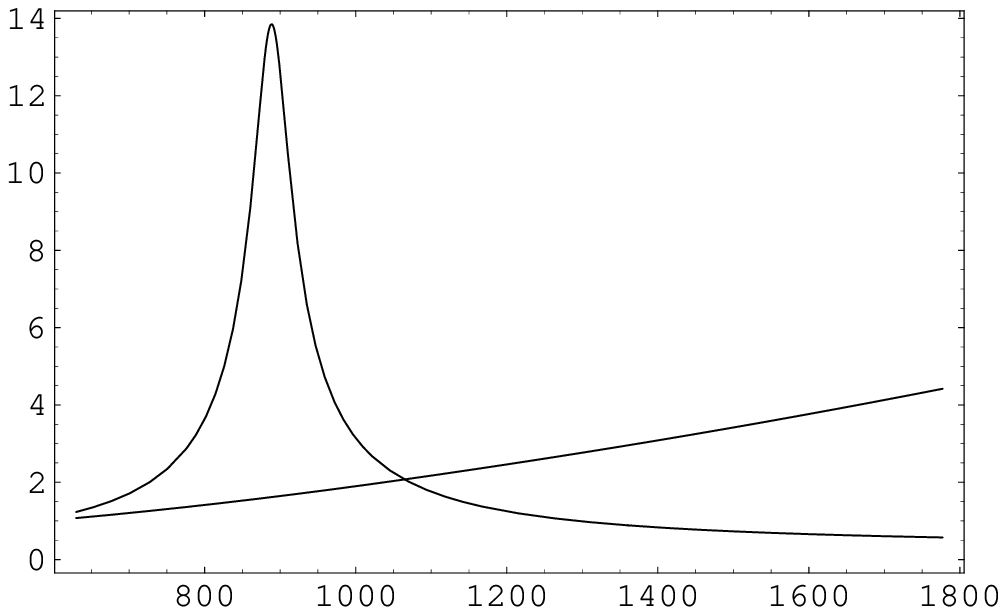}} & 
\resizebox{35mm}{!}{\includegraphics{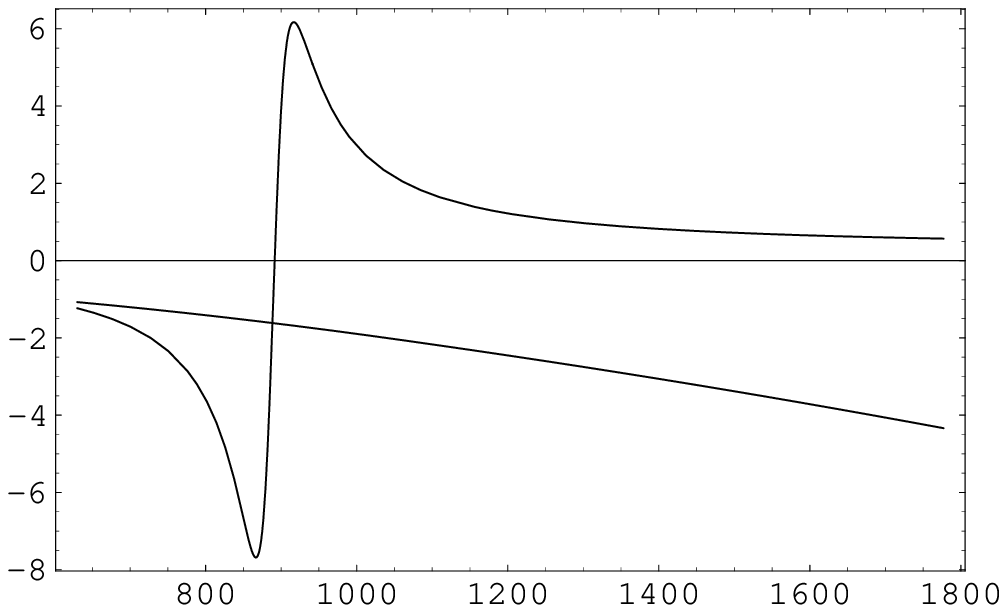}} 
\\
$|F_S|$ & ${\rm Re} F_S$  \\
\resizebox{35mm}{!}{\includegraphics{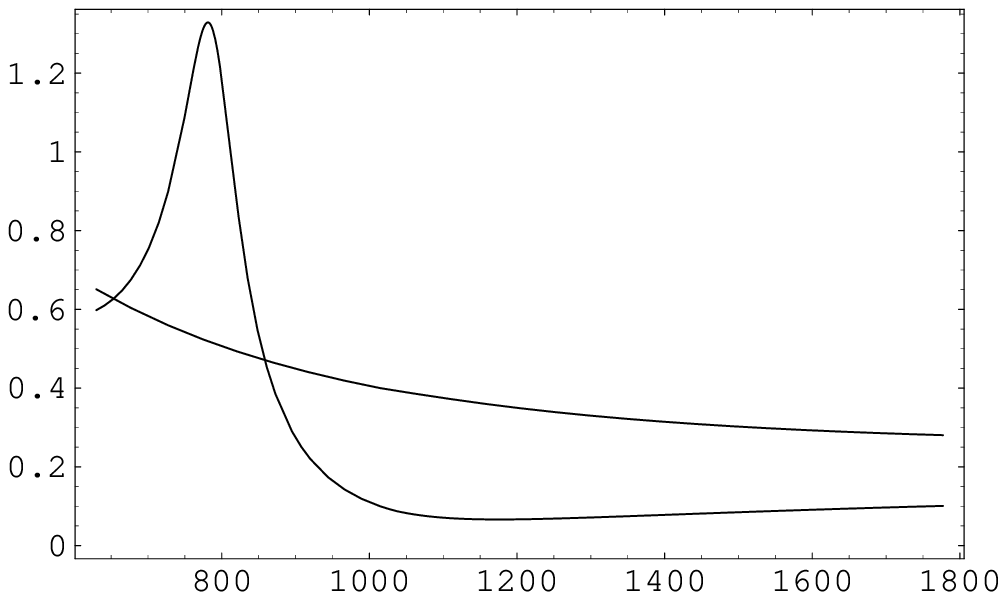}}  & 
\resizebox{35mm}{!}{\includegraphics{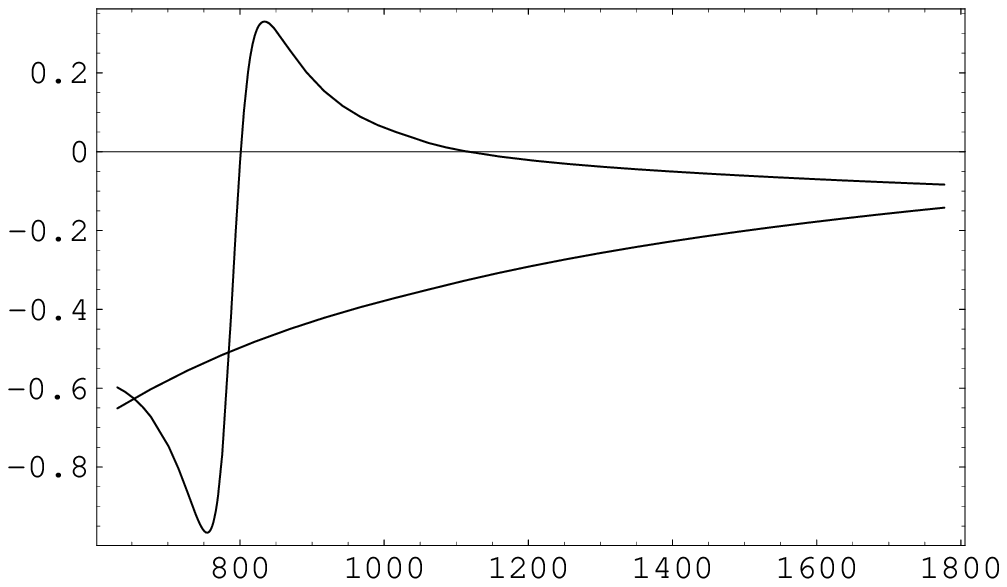}} 
\\
${\rm Im} F $ & ${\rm Im} F_S$ \\
\resizebox{35mm}{!}{\includegraphics{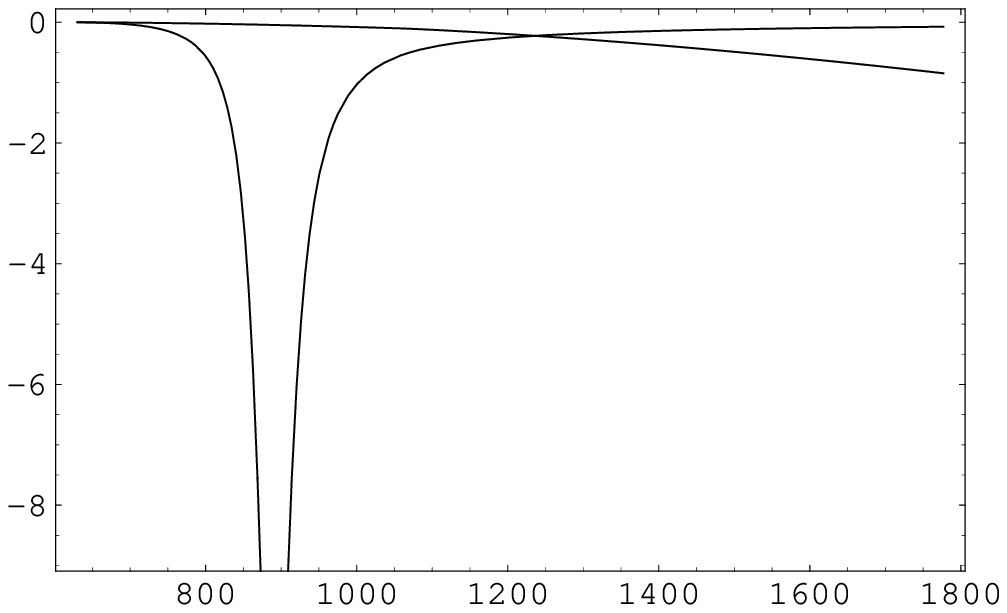}} & 
\resizebox{35mm}{!}{\includegraphics{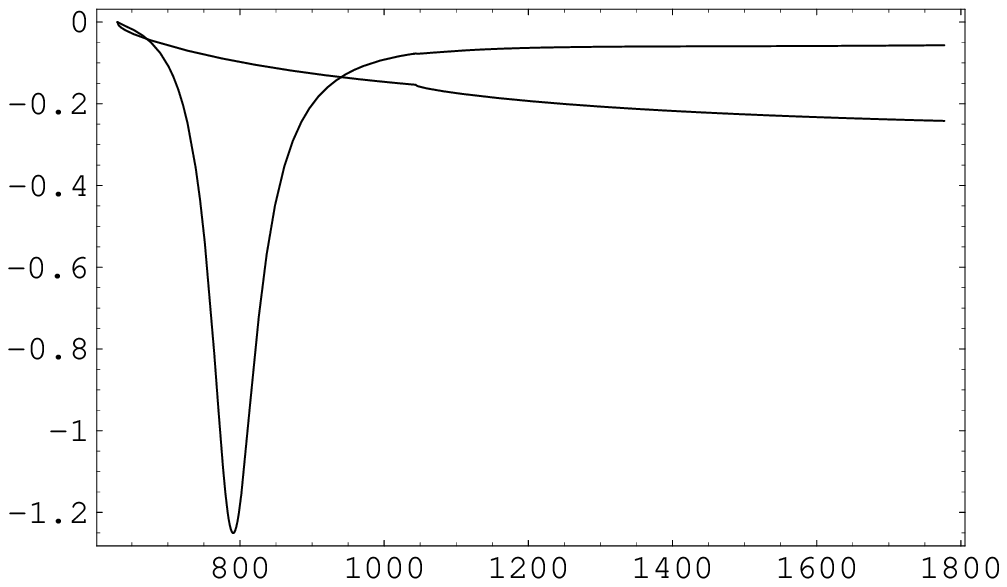}} 
\end{tabular}
\end{center}
\caption{The form factors as functions of $\sqrt{s}$ (MeV)
predicted by the chiral Lagrangian including 
the scalar and the vector resonances Eq.(\ref{lagrangian}).
They are compared with the form factors predicted by chiral
perturbation.}
\label{ffs}
\end{figure}

\begin{figure}
\begin{center}
\resizebox{55mm}{!}{\includegraphics{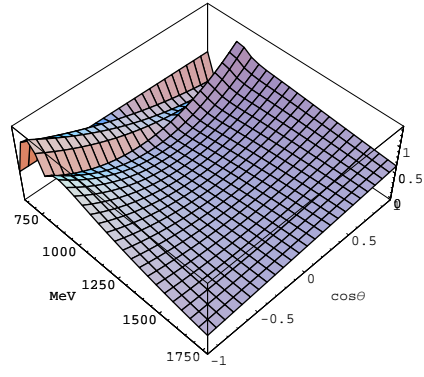}} 
\end{center}
\caption{The vertical axis denotes
the normalized double differential rate 
$\frac{d^2\Gamma}{d \sqrt{s} d \cos \theta}/
\frac{d \Gamma}{d \sqrt{s}}$.}
\label{double}
\end{figure}

\begin{figure}
\begin{center}
\resizebox{55mm}{!}{\includegraphics{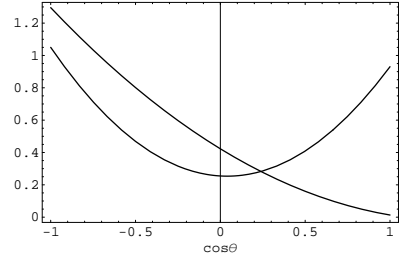}}
\end{center}
\caption{
The same as Fig.(\ref{double}) but with $\cos \theta$ 
distribution for the fixed $\sqrt{s}$ chosen as
$\sqrt{s}=700,900{\rm (MeV)}$. For $\sqrt{s}=900$,
the angular distribution is more asymmetric than
the case for $\sqrt{s}=700$ with respect
to the replacement of $\cos \theta \to -\cos \theta$.}
\label{double2}
\end{figure}

\begin{figure}[htbp]
\begin{center}
\includegraphics[width=8cm]{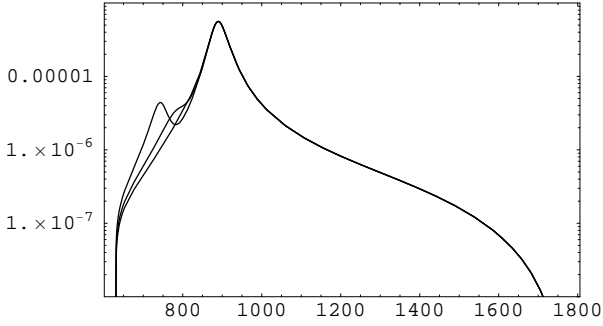}
\end{center}
\caption{The prediction of the
differential branching fraction for $\tau \to K^- \pi^0
\nu$. The holizontal
axis shows the hadronic invariant mass $\sqrt{s}$ (MeV).
The vertical axis shows $\frac{\rm d Br}{d \sqrt{s}}$.
We vary the scalar meson mass as 
$M_{\kappa}=750, 800, 850$. As for comparison with
experimental data, the spectrum for $\tau^{\pm} \to
 K_s \pi^{\pm} \nu $ 
has been measured by Belle \cite{Belle:2007rf}.}
\label{Figlog}
\end{figure}

To predict the forward and the backward asymmetry, we
must estimate the form factors. We have used an effective
chiral Lagrangian including the vector \cite{Bando:1984ej}
and the scalar \cite{Ecker:1988te} 
resonances. 
\def\sq{\sqrt{2}}
\bea
{\cal L}&=&\frac{f^2}{4} {\rm Tr} D U D U^{\dagger}+B 
{\rm Tr}{M (U+U^{\dagger})} \nn \\
&+& 
{\rm Tr}D_{\mu} S D^{\mu} S-M_{\sigma}^2 {\rm Tr}S^2 \nn \\
&-&\frac{1}{2} {\rm Tr} F_{\mu \nu} F^{\mu \nu}
+ M_{\rho}^2 {\rm Tr}(V_{\mu}-\frac{\alpha_{\mu}}{g})^2 \nn \\
&+&\frac{g_1}{4} {\rm Tr}(D_{\mu} U D^{\mu} U^{\dagger}) (\xi S \xi^{\dagger})
\nn \\
&+& g_2 {\rm Tr} \left((\xi M \xi + \xi^{\dagger} M \xi^{\dagger})
S \right),
\label{lagrangian}
\eea
where $U$ is the chiral field defined by,
\bea
U&=& \xi^2, \ \xi=\exp(i\frac{\pi}{f}), \nn \\
\pi&=& \left(\begin{array}{ccc} 
                   \frac{\pi^0}{2}+\frac{\eta_8}{2 \sqrt{3}}
 & \frac{\pi^+}{\sq}& \frac{K^+}{\sq} \\
 \frac{\pi^-}{\sq} & -\frac{\pi^0}{2} +
\frac{\eta_8}{2 \sqrt{3}} & \frac{K^0}{\sq} \\
\frac{K^-}{\sq} & \frac{\overline{K^0}}{\sq} & 
-\frac{\eta_8}{\sqrt{3}} \end{array}
\right).
\eea
$M$ is the chiral breaking term,
\bea
M=\left(\begin{array}{ccc} m_u & 0 & 0 \\
                           0   & m_d & 0 \\
                           0   & 0  & m_s \end{array} \right).
\label{chiral}
\eea
The scalar and the vector fields are given as,
\bea
S&=&\left(\begin{array}{ccc} 
  \frac{\delta^0}{2}+\frac{\sigma}{2}
 & \frac{\delta^+}{\sq}& \frac{\kappa^+}{\sq} \\
 \frac{\delta^-}{\sq} & -\frac{\delta^0}{2} +
\frac{\sigma}{2} & \frac{\kappa^0}{\sq} \\
\frac{\kappa^-}{\sq} & \frac{\overline{\kappa^0}}{\sq} & 
\frac{\sigma_{ss}}{\sq} \end{array}
\right) +S_0  \nn \\
V&=&\left(\begin{array}{ccc} 
  \frac{\rho}{2}+\frac{\omega}{2}
 & \frac{\rho^+}{\sq}& \frac{K^{\ast +}}{\sq} \\
 \frac{\rho^-}{\sq} & -\frac{\rho^0}{2} +
\frac{\omega}{2} & \frac{K^{\ast 0}}{\sq} \\
\frac{K^{\ast -}}{\sq} & \frac{\overline{K^{\ast 0}}}{\sq} & 
\frac{\phi}{\sqrt{2}} \end{array}
\right),
\eea
where $S_0$ is the vaccum expectation value of the scalar 
and is given as
$S_0=\frac{g_2 M}{M_{\sigma}^2}$. 
We note the kinetic terms of the scalar generate the mixing of the
scalar ($\kappa(0^+)$) and the vecor
mesons ($K^{\ast}(1^-)$) because,
\bea
&& {\rm Tr} 
D_{\mu} S D^{\mu} S \nn \\
&=&{\rm Tr}(\partial_{\mu} S +i g [V_{\mu}, S_0])
(\partial^{\mu}S+ig[V^{\mu}, S_0]).\nn \\
\eea
We have computed the hadronic form factors relevant for the processes $
\tau \to \nu K \pi^0$ defined in 
Eq.~(\ref{formfactors}).
The vector form factor is given as,
\bea
F(Q^2)&=&\frac{1}{\sqrt{2}}
\left( -\frac{R+R^{-1}}{2} \right.\nn \\
&& \left. + \frac{M_{\rho}^2}{2 g^2 F_K F_{\pi}}
(1-\frac{M_{\rho}^2}{M_{K^{\ast}}^2-Q^2}) \right).
\label{vectorff}
\eea
The scalar form factor is,
\bea
&& Q^2F_S^{K^+ \pi^0}=-(m_s-m_u) <K^+ \pi^0|\overline{u}s|0>\nn \\
&=& \frac{1}{2 \sqrt{2}}\left(-\Delta_{K \pi} (R+R^{-1})-
\frac{Q^2 M_{\kappa}^2 }{M_{\kappa}^2-Q^2} (R-R^{-1}) \right. \nn \\
&+&\left. \frac{Q^2}{M_{\kappa}^2-Q^2} (\frac{m_{\pi}^2}{2 \Delta}-
\frac{m_K^2}{1+\Delta}) (2 \Delta R +(1+\Delta)R^{-1}) \right.\nn \\
&+&\left. \frac{\Delta_{K \pi}}{M_{\kappa}^2-Q^2}(1-\Delta) 
(-\frac{m_{\pi}^2}{2 \Delta}R^{-1}+
\frac{m_K^2}{1+\Delta} R) \right),
\label{scalarff} 
\eea
where 
$R=\frac{F_K}{F_{\pi}}, 
\Delta_{K \pi}=m_K^2-m_{\pi}^2, \Delta=\frac{m_u+m_d}{2 m_s}
 \sim \frac{1}{25}$.
The form factors are computed using the chiral Lagrangian
of Eq.~(\ref{lagrangian}). 
The major contribution of the $\tau \to \nu K \pi$ decay comes from
the decay chain of $\tau \to K^* \nu \to \nu K \pi$.
We also have the contribution from the scalar resonance $\kappa$
as $\tau \to \kappa \nu \to \nu K \pi$. The latter contributes
to the scalar form factor. 
The form factors given in Eq.~(\ref{vectorff}) and Eq.~(\ref{scalarff})
do not include the contribution of the width of the resonances.
We have included the width by computing the imaginary part
of the self-energy corrections for $K^{\ast}$, $\kappa$ and their
mixing.
Then for instance,
the inverse propagator for $K^{\ast}$ is modified as,
\bea
A&=&M_{K^{\ast}}^2-Q^2-i  M_{K^{\ast}}\Gamma_K^{\ast}(Q^2)\nn \\
\Gamma_{K^{\ast}}(Q^2)&=&\frac{3}{48 \pi M_{K^{\ast}}} 
\left(
\frac{\nu_{K \pi}^3}{Q^4}
+ \frac{\nu_{K \eta}^3}{Q^4} 
 \left(\frac{F_{\pi}}{F_8}\right)^2 \right) g_{K^{\ast} k \pi}^2
,\nn \\
\eea
where
$g_{K^{\ast} K \pi}=
\frac{M_{\rho}^2}{4 g F_K F_{\pi}}=3.246$
which reproduces $
\Gamma_{K^{\ast}}(M_{K^{\ast}}^2)=50.8 ({\rm MeV})
$. The form factors including the width of the $K^*$ and
$\kappa$ are compared with the prediction of the
chiral perturbation
theory in Fig.~(\ref{ffs}).
Using the form factor described above, we
have computed the double differential rate 
in Fig.~(\ref{double}),
the angular distribution in Fig.~(\ref{double2})
and the hadronic invariant mass spectrum
$\frac{\rm d Br}{ d \sqrt{s}}$ in Fig.~(\ref{Figlog}).
The branching fraction for $\tau^{\mp} \to K^{\mp} \pi^0 \nu$
is obtained by integrating the hadronic invariant mass spectrum.
The theoretical prediction is given
by,
\bea
{\rm Br}(\tau^+ \to \pi^0 K^+ \nu)_{\rm th}=
0.448^{+0.004}_{-0.003} \%,
\label{Brth}
\eea where we have changed the scalar meson mass
as $M_\kappa=800\mp50$. The theoretical prediction
of Eq.~(\ref{Brth})
can be compared with 
the experimental result
from Babar~\cite{Aubert:2007jh}.
\bea
{\rm Br}=0.416 \pm 0.003\pm 0.018 \%.
\label{Brex}
\eea
One may also use the branching fraction
$\tau^{\pm} \to K_s \pi^{\pm} \nu$ to estimate 
the $\tau^{\pm} \to K^{\pm} \pi^0 \nu$ in the isospin
limit,
\bea
{\rm Br}(\tau^{\pm} \to K_s^{\pm} \pi \nu)&=&
|p|^2 Br(\tau^{+} \to K^0 \pi^+ \bar{\nu}) \nn \\
&+&|q|^2 Br(\tau^- \to
\bar{K^0} \pi^-  \nu), \nn \\
&=&(1+|\epsilon_m|^2) {\rm Br}(\tau^{-} \to \bar{K^0} \pi^- \nu),\nn \\
&=& 2 (1+ |\epsilon_m|^2) {\rm Br}(\tau^- \to K^- \pi^0 \nu) \nn \\
&\simeq & 0.90 \%,
\label{Ks} 
\eea
where 
$p=\frac{1+\epsilon_m}{\sqrt{2}}, q=\frac{1-\epsilon_m}{\sqrt{2}}$.
The numerical value of 
Eq.(\ref{Ks}) should be compared with the Belle measurement
\cite{Belle:2007rf}
${\rm Br}(\tau \to K_s \pi \nu)=0.808 \pm 0.004 \pm
0.026 \%$.
In Eq.(\ref{Ks}), we assumed 
$\frac{{\rm Br}(\tau^- \to \bar{K^0} \pi^- \nu)}
{{\rm Br}(\tau^+ \to K^0 \pi^+ \bar{\nu})}=1$
and use the isospin relation
$
{\rm Br}(\tau^- \to K^0 \pi^- \nu)=2 
{\rm Br}(\tau^- \to K^- \pi^0 \nu).
$
\section{New Physics interaction and source of CP violation}
\begin{figure}[htbp]
\begin{center}
\includegraphics{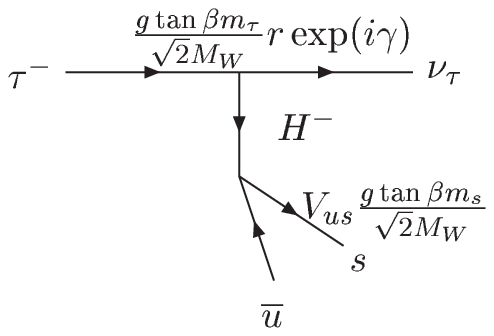}
\end{center}
\caption{The CP violating charged Higgs contribution 
to $\tau \to K \pi \nu_{\tau}$ decay.}
\label{feynchargedhiggs}
\end{figure}

Having explored the various distributions
of the decay within the standard model, we
turn to CP violation of the two Higgs doublet model.
It has been known that the two Higgs doublet model
with the condition of the natural flavor conservation,
the charged Higgs coupling to the $\tau$ lepton and neutrino
is real as in the standard charged current interaction.
Therefore within the scheme, we may not have the CP
phase. Then we relax the condition of natural flavor
condition as \cite{Kimura:2008gh},
\bea
-{\cal L}&=&y_{1 ij} \overline{e_{Ri}} \tilde{H_1}^{\dagger}
l_{Lj}+ \nn \\
&+& y_{2 ij} \overline{e_{Ri}} H_2^{\dagger} l_{Lj} 
+ y^{\nu}_{2i} \overline{\nu_{Ri}} \tilde{H_2}^{\dagger} l_{Li}
+ {\rm h.c.}.
\eea
We allow the charged lepton to acquire mass from two Higgs 
$H_1$ and $H_2$. One can, in general, take the following 
parametrizations for Higgs fields,
\bea
H_1&=&e^{i \frac{\theta_{CP}}{2}}\left( \begin{array}{c}
\frac{v_1+h_1-i \sin \beta A}{\sqrt{2}} \\
-\sin \beta H^- \\
\end{array} \right), \nn \\
H_2&=&e^{i \frac{\theta_{CP}}{2}}\left( \begin{array}{c}
-\cos \beta H^+ \\
\frac{v_2+h_2-i \cos \beta A}{\sqrt{2}} \\
\end{array} \right),
\eea
where
the relative phase of the vaccuum expectation value of the
two Higgs is denoted by
$\theta_{CP}$. 
In this case, the CP violation occurs in the charged Higgs
coupling with $\tau$ lepton and neutrino.
One can parametrize the coupling \cite{Kimura:2008gh},
\bea
{\cal L}&=&H^+ \overline{\nu_{L i}} l_{Rj}
(\frac{Y_{2 ji}^{\ast} e^{+i \frac{\theta_{CP}}{2}}}{\cos \beta}
-\delta_{ij} \frac{g \tan \beta m_{j} } {\sqrt{2} M_W}) \nn \\
\eea
where $Y_2=V_R y_2 V_L^{\dagger}$ and $V_R$ and $V_L$
are unitary matices which diagonalize the charged lepton mass
matrix.
Then one can show that the charged Higgs coupling to $\tau$
lepton and neutrino can be complex and CP violating as shown
in Fig.~\ref{feynchargedhiggs}.
We have introduced the following parametrization,
\bea
r \exp(i \gamma)=1-
\frac{\sqrt{2} M_W}{m_{\tau}} \frac{Y_{2 \tau \tau}^{\ast} 
e^{i \frac{\theta_{CP}}{2}}}{g \sin \beta}.
\eea
In the minimal two Higgs doublet model with the natural
flavor conservation, $r=1$ and $\gamma=0$. Note that
CP phase may come from either CP violating phase $\theta_{CP}$
of Higgs vaccum expectation value or 
the new Yukawa couplings
$y_2$. 
The charged Higgs contribution can be easily
incorporated by replacing the scalar form factor
$F_S$ as,
\bea
F_S \rightarrow 
F_S(1-\frac{Q^2}{M_H^2} \tan^2 \beta r \exp(\pm i\gamma)),
\eea 
for $\tau^{\mp}$ decay.
\begin{figure}
\begin{center}
\includegraphics[width=7cm]{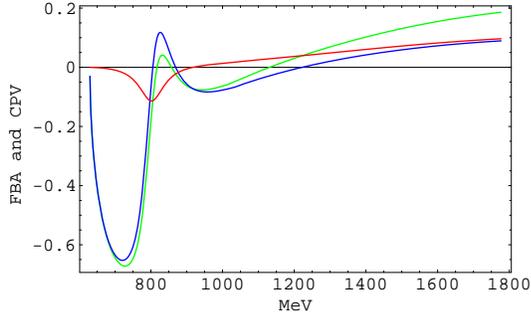} 
\end{center}
\caption{The forward and backward asymmetries for
$\tau^-$ decay (green) and $\tau^+$ decay (blue).
CP violation of the
forward and backward asymmetries defined in Eq.(\ref{CPV})
is shown in the red
solid line. We choose, 
$M_H=200$(GeV), $\tan \beta=50, r=2$,$\gamma=\frac{\pi}{2}$.
$M_{\kappa}=800$(MeV).}
\label{CPVf}
\end{figure}
We define the CP violation of the forward and the
backward asymmetry,
\bea
A_{FB}(s)-\bar{A}_{FB}(s).
\label{CPV}
\eea
In Fig.(\ref{CPVf}), we have plotted  
the forward
and backward asymmetries  and their 
CP violation in Eq.(\ref{CPV}).
We have shown the dependence of the
CP violation of the forward and the backward asymmetry
on the CP violating phase $\gamma$ as 
$\frac{\pi}{2}, \frac{\pi}{6},
$ and $\frac{\pi}{18}$ in Fig.(\ref{Phasef}).
We also have changed $\kappa$ mass and studied how the 
CP violation depends on the resonance parameter in 
Fig.(\ref{kappamass}).

\begin{figure}
\begin{center}
\includegraphics[width=7cm]{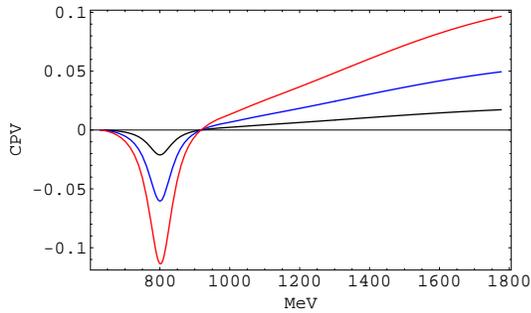} 
\end{center}
\caption{CP phase dependence ($\gamma=\frac{\pi}{2}, \frac{\pi}{6}
,\frac{\pi}{18} $) of the CP violation of the forward
and the backward asymmetry.}
\label{Phasef}
\end{figure}
\begin{figure}
\begin{center}
\includegraphics[width=7cm]{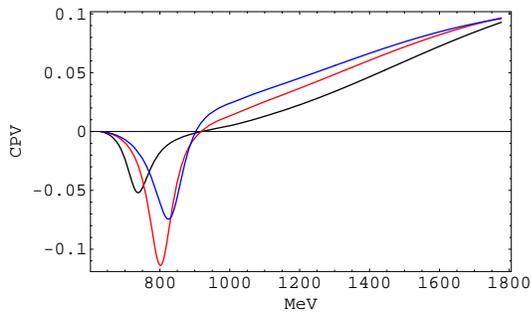} 
\end{center}
\caption{The dependence on the mass of $\kappa$ 
($M_{\kappa}=700,800,850$(MeV)) of the CP violation 
of the forward and backward asymmetry.}
\label{kappamass}
\end{figure}
\section{Summary}
\begin{itemize}
\item We have presented a realistic calculation of the
form factors of the $\tau$ decays. 
We study the vector and scalar form factors including 
$\kappa(800)$ and $K^{\ast}$. The hadron invariant 
mass spectrum is obtained. 
\item We study the forward and backward asymmetry and find
the large asymmetry (not CP) $\sim 60 \%$ near the threshold region
within the standard model using the form factors which we obtain.
\item Including the new physics contribution from charged
Higgs exchange, we have seen that CP violation of the
forward and backward asymmetry can be as large
as $10 \%$. ($M_H=200, r=2, \tan \beta=50,\gamma=\frac{\pi}{2}$)
\item{The CP violating source of the charged Higgs coupling
 is identified as the non-minimal 
($r\ne 1, \gamma \ne 0$)
two Higgs doublet model structure. We have shown the CP violation
may originate from the phase of the vaccuum expectation 
value of Higgs field.}
\end{itemize}
\begin{acknowledgments}
This work was supported by KAKENHI, Grand in Aid for
Scientific Research on Priority Areas 
" New Development for Flavor Physics",
No.19034008 and No.20039008 of MEXT, Japan.
The work of K. Y. L was supported by 
WCU program through the KOSEF funded by the MEST (R31-2008-000-10057-0) and by the Korea Research Foundation Grant funded by the Korean Government(KRF-2008-313-C00167).
\end{acknowledgments}
\bigskip 
\def\apj#1#2#3{Astrophys.\ J.\ {\bf #1}, #2 (#3)}
\def\ijmp#1#2#3{Int.\ J.\ Mod.\ Phys.\ {\bf #1}, #2 (#3)}
\def\mpl#1#2#3{Mod.\ Phys.\ Lett.\ {\bf A#1}, #2 (#3)}
\def\nat#1#2#3{Nature\ {\bf #1}, #2 (#3)}
\def\npb#1#2#3{Nucl.\ Phys.\ {\bf B#1}, #2 (#3)}
\def\npps#1#2#3{Nucl.\ Phys.\ Proc. \ Suppl. {\bf #1C}, #2 (#3)}
\def\plb#1#2#3{Phys.\ Lett.\ {\bf B#1}, #2 (#3)}
\def\prd#1#2#3{Phys.\ Rev.\ {\bf D#1}, #2 (#3)}
\def\ptp#1#2#3{Prog.\ Theor.\ Phys.\ {\bf #1}, #2 (#3)}
\def\pr#1#2#3{Phys.\ Rev.\ {\bf #1}, #2 (#3)}
\def\prl#1#2#3{Phys.\ Rev.\ Lett.\ {\bf #1}, #2 (#3)}
\def\prp#1#2#3{Phys.\ Rep.\ {\bf #1}, #2 (#3)}
\def\sjnp#1#2#3{Sov.\ J.\ Nucl.\ Phys.\ {\bf #1}, #2 (#3)}
\def\zp#1#2#3{Z.\ Phys.\ {\bf #1}, #2 (#3)}
\def\jhep#1#2#3{JHEP\ {\bf #1}, #2 (#3)}
\def\epjc#1#2#3{Eur. Phys. J.\ {\bf C#1}, #2 (#3)}
\def\rmp#1#2#3{Rev. Mod. Phys.\ {\bf #1}, #2 (#3)}
\def\prgth#1#2#3{Prog. Theor. Phys.\ {\bf #1}, #2 (#3)}

\begin{thebibliography}{99} 
\bibitem{Bigi:2005ts}
I. I. Bigi and A. I. Sanda,
\plb{625}{47}{2005}.
\bibitem{Calderon}
G. Calderon, D. Delepine and G.Lopez Castro
\prd{75}{076001}{2007}.
\bibitem{Bonvicini:2001xz} G. Bonvicini et al. (CLEO),
\prl{88}{111803}{2002}.
\bibitem{Kimura:2008gh}
D. Kimura, K. Y.  Lee, T. Morozumi,
                  and K. Nakagawa,
arXiv:0808.0674[hep-ph].
\bibitem{Kuhn:1996dv}J. H. Kuhn. and E. Mirkes,
\plb{398}{407}{1997}.
\bibitem{Tsai:1996ps}Y. S. Tsai,
\npps{55}{293}{1997}.
\bibitem{Choi:1998yx} S. Y. Choi, J. Lee, and J. Song,
\plb{437}{191}{1998}.
\bibitem{Beldjoudi:1994hi}
L. Beldjoudi and T. N. Truong,
\plb{351}{357}{1995}.
\bibitem{Bando:1984ej}
M. Bando, T. Kugo, S. Uehara, K. Yamawaki and T. Yanagida,
\prl{54}{1215}{1985}.
\bibitem{Ecker:1988te} G. Ecker and J. Gasser and A. Pich and E. de Rafael,
\npb{321}{311}{1989}.
%
\bibitem{Belle:2007rf} D. Epifanov et al.(Belle),
\plb{654}{65}{2007}.
\bibitem{Aubert:2007jh} B. Aubert et al.(Babar),
\prd{76}{051104}{2007}.
%
\end{thebibliography}

\end{document}